\documentclass[conference]{IEEEtran}
\IEEEoverridecommandlockouts
\usepackage{cite}
\usepackage{amsmath,amssymb,amsfonts}
\usepackage{algorithmic}
\usepackage{graphicx}
\usepackage{textcomp}
\usepackage{xcolor}
\usepackage{hyperref}
\def\BibTeX{{\rm B\kern-.05em{\sc i\kern-.025em b}\kern-.08em
    T\kern-.1667em\lower.7ex\hbox{E}\kern-.125emX}}
\usepackage[font=normalsize,labelfont=bf]{caption}
\begin{document}

\title{Cloud-Based Deep Learning: End-To-End Full-Stack Handwritten Digit Recognition}

\author{\IEEEauthorblockN{Ruida Zeng, Aadarsh Jha, Ashwin Kumar, Terry Luo}
\{ruida.zeng, aadarsh.jha, ashwin.kumar, terry.luo\}@vanderbilt.edu\\
\IEEEauthorblockA{School of Engineering\\
Vanderbilt University, Nashville, Tennessee, USA \\
}}

\maketitle
\begin{abstract}
Herein, we present \textit{Stratus}, an end-to-end full-stack deep learning application deployed on the cloud. The rise of productionized deep learning necessitates infrastructure in the cloud~\cite{b1} that can provide such service (IaaS). In this paper, we explore the use of modern cloud infrastructure and micro-services to deliver accurate and high-speed predictions to an end-user, using a Deep Neural Network (DNN) to predict handwritten digit input, interfaced via a full-stack application. We survey tooling from Spark ML, Apache Kafka, Chameleon Cloud, Ansible, Vagrant, Python Flask, Docker, and Kubernetes in order to realize this machine learning pipeline. Through our cloud-based approach, we are able to demonstrate benchmark performance on the MNIST dataset with a deep learning model.
\end{abstract}


\section{Introduction}
In recent years, Deep Neural Networks and Deep Learning techniques have been leveraged to varying tasks due to their efficiency, accuracy, and ease-of-implementation~\cite{b2}. In particular, the task focused on within this project is classification – we apply deep learning methods to predict  handwritten digits from the end-user. In order to train our model, we utilize the MNIST~\cite{b3} data set of handwritten digits. 

A simple machine learning pipeline involves feeding input to a trained model, and returning the predicted information. However, in modern production environments, such a naive approach will not scale to user demand and changes in input data~\cite{b4}. For this reason, concepts of cloud computing, such as distributed workload, load balancing, cloud-as-a-service (CaaS), dockerization, container orchestration, large-scale data processing, as well as full automation are explored in order to create a productionized machine learning pipeline. Working in tandem, both deep Learning and cloud Computing principles may create pipelines that are resilient to large-scale and variable user input. 

\begin{figure*}[htbp]
    \centering
    \includegraphics[width=\textwidth]{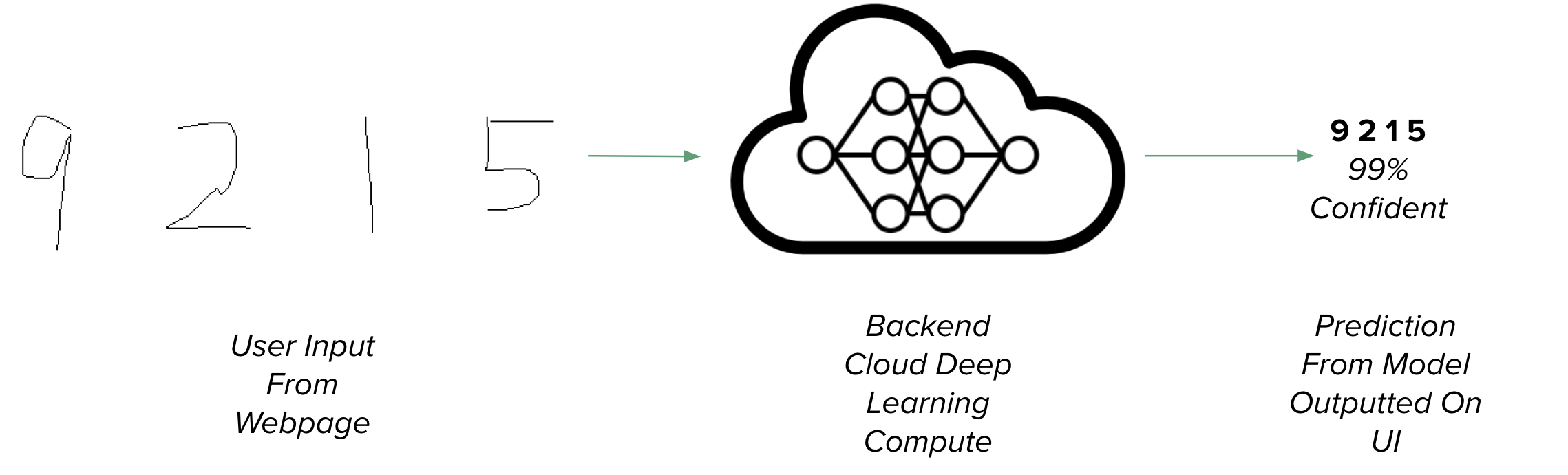}
    \caption{An Overview of the Interaction between the Cloud-Based and Web-Based Infrastructure.}
    \label{overview}
\end{figure*}

The primary motivation to build \textit{Stratus} is to combine the salient concepts of deep learning, cloud computing, and web architecture in order to create an end-to-end pipeline by which users may enter handwritten digits, and receive an outputted value (Figure \ref{overview}). In effect, this project is a method of replicating a productionized, modern environment in deploying a deep learning pipeline, but in a smaller, more accessible scale. In particular, our deep learning pipeline will allow for model training, prediction, and output. Our cloud pipeline will orchestrate the deep learning infrastructure in a persistent, distributed, and self-contained manner. Our full-stack component will then interface with the user and our deep learning and cloud backend, allowing for ease-of-use and an intuitive user interface. 

Furthermore, we would like to primarily take advantage of of the Spark architecture we have built so far in a previous project, as well as the automated tooling in previous programming assignments. The complexity of the project necessitates four people as we are combining ideas of the cloud, distributed workloads, full-stack development, and machine/deep learning core concepts. Our goal is to match the industry way of deploying and utilizing machine learning (ML) models in the cloud by practicing using a productionized data pipeline system for real-time processing. Overall, our plan is to combine and utilize methods learned in this class for this capstone project.  

\label{Design, Architecture, and Implementation}
\section{Design, Architecture, and Implementation}

Herein, we enumerate the design, architecture, and implementation details of the three salient portions of our project, including the: 1) cloud-based, 2) full-stack-based, and 3) deep-learning-based components of our project.

\begin{figure*}[htbp]
    \centering
    \includegraphics[width=\textwidth]{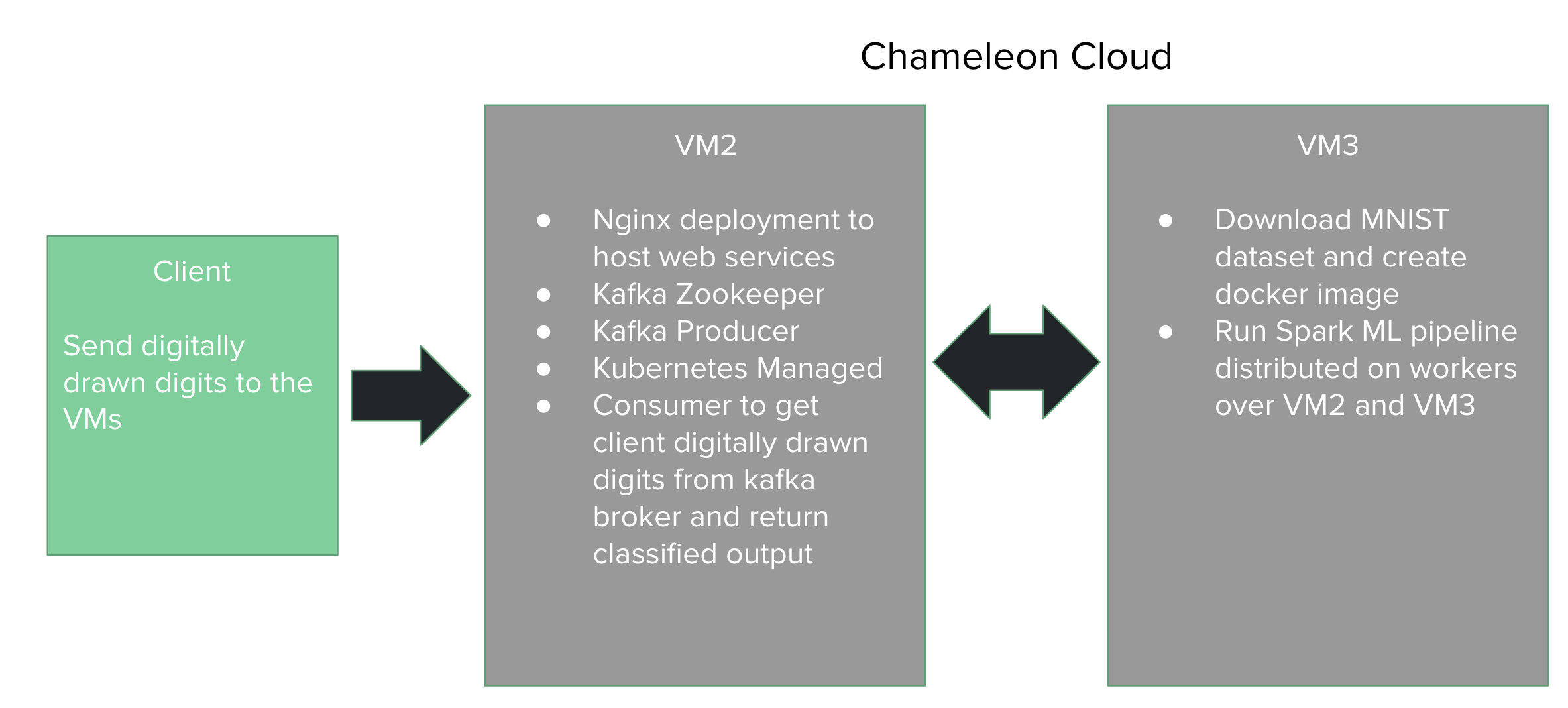}
    \caption{An Overview of the Cloud Infrastructure Partitioned Across Both VMs.}
    \label{architecture}
\end{figure*}

\subsection{Cloud-Based Implementation Details}

Our cloud based architecture aims to advance the software we have previously built in our programming assignments (Figure \ref{architecture}). Our main goal is to digitally draw digits on the client interface and return the digit drawn to the client by deploying several Docker containers to handle the cloud architecture. All of the Docker containers we create will be managed through Kubernetes and balanced over two VMs, specifically VM2 and VM3. Both VMs will be launched using IaaS software such as Ansible and Vagrant. In fact, Vagrant will act as the VM, not on the client machine, but on our machine to run the appropriate Ansible playbook and deploy the necessary service. We will use a main master playbook to handle the operations and will maintain IP-agnostic functionality. This way, we do not need to consider the availability of floating IP addresses in Chameleon KVMs. We will further break down our master playbook into the following sections: 1) local VM package installation; 2) VM creation; 3) firewall setting configuration; 4) cloud package installation; 5) Docker and Kubernetes installation; 6) kubemaster assignment; 7) Kubeworker2 assignment; 8) Kafka/ZooKeeper image creation and deployment; 9) CouchDB image creation and deployment; 10) Spark image creation and deployment; 11) consumer image creation and deployment; 12) NGINX image creation and deployment; and 13) Spark image creation and deployment.

We intend to create VM2 as an ml.large instance and VM3 as an ml.medium instance to balance load while also considering necessary usage for the Chameleon compute resources. The firewall configuration settings will match the security ports allocated in our security along with additional functionality. Docker and Kubernetes will be installed such that the master is tainted and Docker can run as a daemon. 

The client will access the web-interface, our front-end code, which will be hosted and load-balanced using NGINX (port 30009). We will maintain three NGINX replicas, which will be managed by Kubernetes. Our client will not interact with the Kafka/ZooKeeper brokers or consumer code; instead, for our purposes, we will create a backend that takes in the information sent by the front-end. Specifically, we will deploy a Flask backend (port 30005) in a Docker container using the latest Ubuntu image so as to serve the necessary functionality through an API request. The back-end will then send the information to a randomly assignment Kafka broker. We will maintain three Kafka brokers (ports 30001-30003) and one zookeeper node (port 30000) to balance the Kafka brokers. Once the backend sends the information to the Kafka broker, then the consumer will process that information. The consumer will run as a job hosted on a docker container managed by Kubernetes. The consumer will then use our deep learning model that we already trained in a distributed fashion using Spark to classify the input array and send a probability array to CouchDB (port 30006). The backend Flask container will then be able to take the CouchDB information, construct a Pandas DataFrame, search for the appropriate array key, and send the answer and probability array to the NGINX hosted front-end. The client will then have the answer as a prediction and a graph listing the probabilities. This combination allows our system to be scalable while maintaining high reliability and durability. 

\subsection{Full-Stack Implementation Details}

\begin{figure}[htbp]
\centerline{\includegraphics[width=80mm]{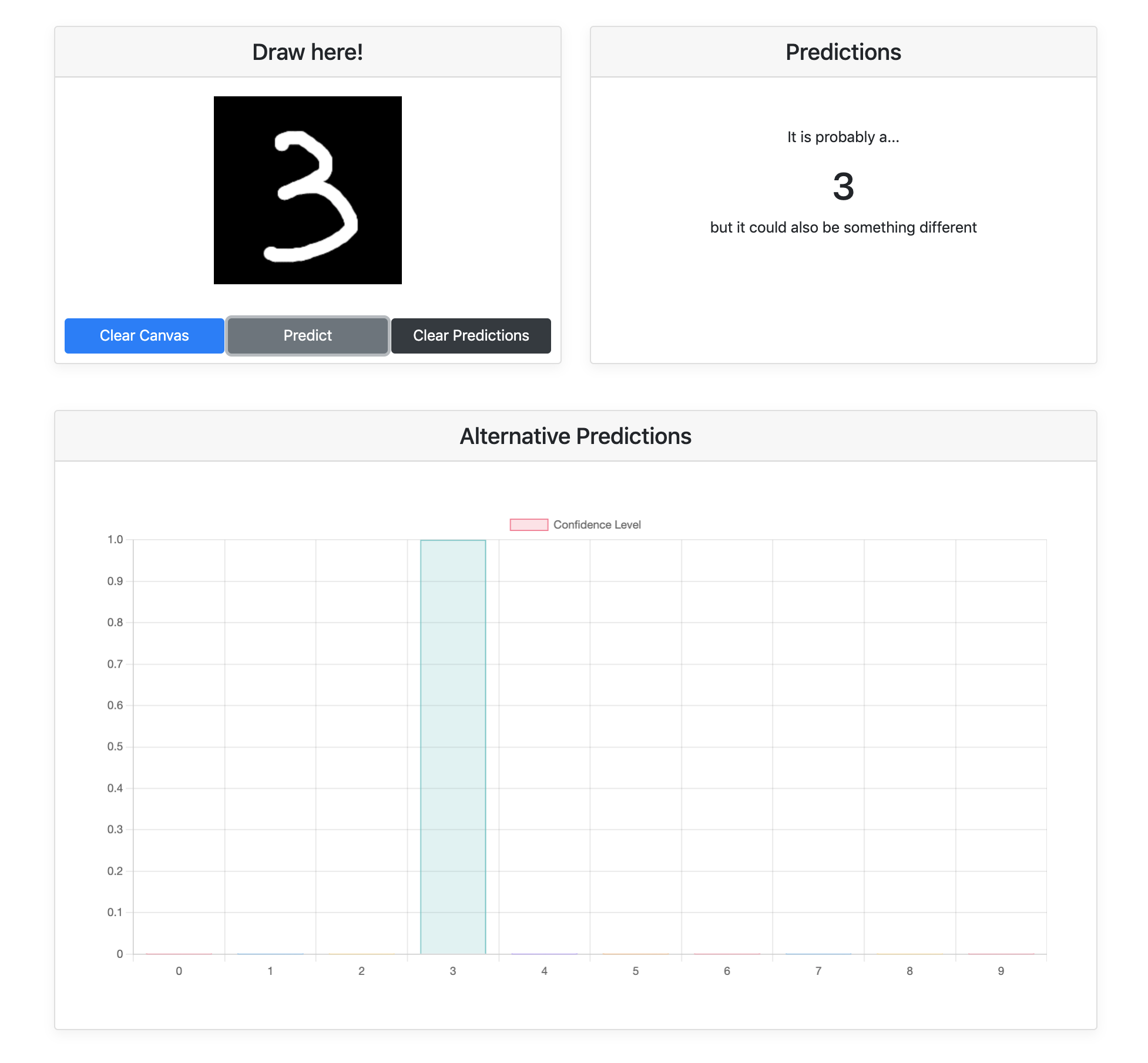}}
\caption{What the User Should Expect after Drawing a Three.}
\label{fig}
\end{figure}

Our project distributes the full-stack nature of our web application into a front-end and a back-end. In the front-end, we utilized JavaScript to dynamically render HTML content based on the results of the cloud and deep learning pipeline. In particular, JavaScript and this back-end/front-end architectural support the overall UI/UX experience that the user is exposed to, as well as communicates with our cloud-based deployment of our deep learning model. 

The back end runs on Flask, which was chosen for ease of use due to its implementation in Python, which is also used as the primary language to support our cloud infrastructure. Our backend's main role is simply to supply an API endpoint for the frontend to make prediction. It provides a POST request that returns digit inferences upon receiving the image data. 

The features of our application (Figure 3) are enumerated below: 
\begin{itemize}
    \item Links to the referenced codebase as well as our deployed codebase on GitHub. The link to the GitHub repository and details regarding the repository are included in Appendix A.
    \item A small canvas, wherein the user can draw the digit they would like to test the model with. In particular, there are three main buttons: 1) \textbf{Clear Canvas}, which clears the drawing canvas; 2) \textbf{Predict}, which allows the user to send their data as a POST request to our backend, which further processes the data via our cloud-based deep learning pipeline; 3) \textbf{Clear Predictions}, which clears the buffer of predicted numbers. 
    \item A prediction pane, which shows the buffer of numbers that the user has requested to be predicted. 
    \item A prediction graph, which shows the ranges of options of numbers (0 - 9), and the associated probability that the model outputted for each number. Of course, the number associated with the maximum probability is logged in the prediction pane.
\end{itemize}

\subsection{Deep Learning Implementation Details}
To implement the deep learning training pipelines, we used Spark, which is known for its parallel computing properties. Since our project aims to explore training a neural network in a distributed fashion, Spark was a natural choice. Specifically, we took advantage of Spark's ML libraries by using the wrapper Python packages such as Elephas. We spent much time understanding the Pythonic version of the implemented libraries as we wanted to take advantage of TensorFlow and Keras, which are easily distributed as Python packages.

In the training phase of our model, we utilize both PySpark and Keras to create and train a deep neural network. The input data is 28 by 28 pixel images of handwritten digits, which is flattened and scaled down to a value between 0 and 1. In particular, in order to facilitate the process of model inference for our web application, we first define the model that is trained. In particular, we use the default MNIST dataset from Keras. A model network is created with Keras as well, with the following layers: 1) a Conv2D layer, 2) a MaxPooling2D layer, 3) a Flatten layer, and 4) two Dense layers. Leveraging this model, we are then able to train on the aforementioned MNIST dataset, which is of 60,000 images, 10\% of which is used to validate our model. The hyperparameters of our network include: a batch size of 64 and 10 epochs. We utilize spark to train our model over 5 distinct workers. We computed the average training time, 144.155361 seconds, over 10 times training the model. 

Then, we evaluate the effectiveness of our trained model using the MNIST test databaset. In particular, the trained model, on a testbed of 10,000 images. Over 10 time average of our trained model, we computed an average accuracy of 0.974500 on our test dataset. It is important to note that the MNIST dataset included data that were handwritten; however, we are trying to classify digitally drawn digits, which may be harder to draw the correct digit. Therefore, it is expected that the performance of our implemented would not be as good as the test dataset results.

\section{Evaluation}
To evaluate the efficacy of our pipeline, we will analyze the following three separate categories holistically: 1) the training time and accuracy of our model in theory; 2) the quality of load balancing for website hosting; and 3) the quality of load balancing for the prediction algorithm. The methods and details of each evaluation categories are explained below, although due to time, scope, and cost constraints, the evaluations only give an rough idea of the outlooks of the project, and more thorough testings could potentially be employed in the future to improve the quality and resilience of the pipeline.

\subsection{Training Time and Accuracy}

Since the focus of the project is to test accuracy of predicting handwritten digits, we decide to test the accuracy of the prediction by simply writing digits manually and recording the results. Essentially, we manually tested 100 digits, from 0 through 9, with each digit being written ten times, and we record the prediction accuracy for all 100 digits.

\begin{figure}[htbp]
\centerline{\includegraphics[width=80mm]{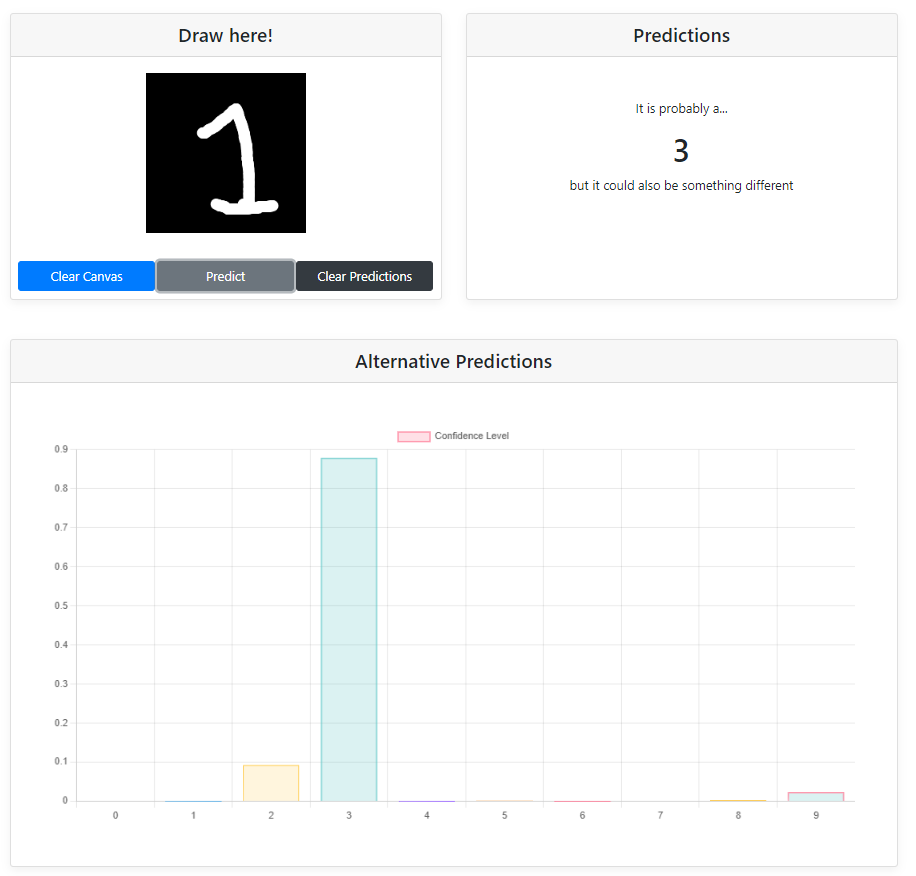}}
\caption{An Example of an Incorrect Prediction.}
\label{fig}
\end{figure}

Overall, it seems like our model is quite accurate on some digits, while being less accurate on other digits (Figure 4). Note that for this evaluation, we only consider the primary prediction and do not take into account the confidence interval. Below is a graph of the accuracy (Figure 5) of each digit after 10 handwritten attempts each. 

\begin{figure}[htbp]
\centerline{\includegraphics[width=80mm]{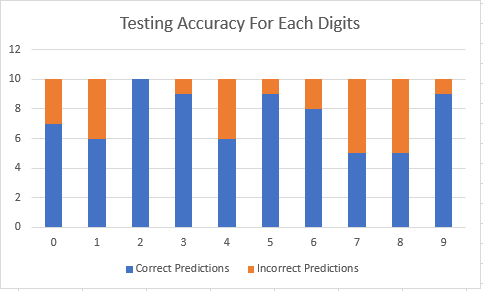}}
\caption{Manual Testing Accuracy for Each Digit.}
\label{fig}
\end{figure}

Based on the manual testing, we can see that the testing accuracy is very good for 2, at 100\% after 10 trials, and fairly good for 3 and 5, both at 90\%. On the other hand, the testing accuracy is relatively low for 7 and 8 at 50\%. The overall testing accuracy for the model is at 74\%. This overall testing accuracy meets the satisfactory threshold for the proof of concept project. However, there is most definitely room for improvement in terms of the machine learning model for handwritten digit recognition and perhaps even the training data set for future applications that require higher digit recognition accuracy.

\subsection{Load Balancing for Website Hosting}

In order to test the load balancing for website hosting, we used a tool called locust, it is an open source Python library that is commonly used for load testing. Not only does it have the ability of implement complex user flows, it also has a web interface that allows visualization of the results as charts with important information.

To test website hosting, we first defined the number of users (peak concurrency) as 50 users (peak concurrency), with a spawn rate (users added/stopped per second) of 5. Then we swarm the website with an arbitrary GET request. Clearly, this is too much traffic for our websites and most requests resulted in an HTTPError (429 Client Error: Too Many Requests for url: http://stratus-final/). Using the 50 users, we issued a total of approximately 10,000 requests for the website, and has a failure rate of about 98\% with an average response time of 306ms. The visual graphs generated by the locust library is shown below (Figure 6-8 in APPENDIX B) for these parameters.

The average response time is very good, although this is likely due to HTTPError being easier to forward as the failure rate he failure rate is unacceptably high for these paremeters. This is not desirable since a fast response time containing an error message as opposed to the actual content does not indicate successful and meaningful user interaction with the website interface.

We then lowered both the user count and spawn rate to 25 users and 3 respectively. After issuing approximately 1000 GET requests, we have a failure rate of 3\% and an average response time of 7123ms. The visual representation for 25 users is shown below (Figure 9-11 in Appendix B).

The response time is significantly longer, but the failure rate is much lower, meaning the GET requests are mostly successfully and the website actually loaded for a majority of the generated users. The longer response time also more accurately depict the expected amount of response time needed for the interaction with the interface to successfully complete with no errors.

Last but not least, we tried a user count of 10 with a spawn rate of 1. We issued a total of 1250 requests.  These parameters yielded the best results so far with a failure rate of approximately 0\% (only 3 failures out of 1250 requests) and an average response time of 2950ms. The visualization is shown below (Figure 12-14 in Appendix B).

These parameters seems to reflect the best operating conditions for our demo server using our current server resources with load balancing implemented for website hosting.

Overall, the results so far have shown that our load balancing implementation can easily handle 10 users simultaneously with relatively fast response times with minimal failures. For 25 users, occasional failures do occur, and the response times are much slower, although a failure of 3\%  is acceptable in our opinions. For 50 users, the speed and volume of the requests proved to be too much to handle for our server. Anything beyond 50 users will likely show the same result of mostly failures.

\subsection{Load Balancing for Prediction Algorithm}

Our next step is to verify the load balancing robustness of our implementation. Since we have already conducted accuracy testing for the machine learning model, we decided that we do not actually need a reliable training data set as part of the POST request tests to the server. Instead, we issued POST requests using a dummy array filled with meaningless numbers that supposedly represent the pixels in the drawing. The array has a length of 784 and normally it was converted from the 28x28 drawing using the canvas using our front end code.

Since we have already established that the website cannot handle 50 users simultaneously as shown in section III. (b), we will only be doing POST request load balance testing for 25 users and 10 users. We issued an approximately 4300 requests, and we have an average failure rate of only 1\% and an average response time of 7412 ms. Note that we decided to issue more requests than originally planned since we were surprised to see that the failure rate for POST request swarming is lower than GET request swarming.

Note that we kept the swarming rate the same at 3 for 25 users, and 1 for 10 users. As far as we know, this should not have a big impact on the results as seen by the graphic representation (Figure 15-20 in Appendix B).

For 10 users, the results are exceptional, as our load balancing could handle 10 users at a time using limited resources and run machine learning algorithm on the thousands of POST requests sent. We issued about 1000 requests, and we have less than 1\% failure rate, and an average response time of 3040ms. 

For the 1\% failure that did occur, most of them occurred during a particular interval, possibly due to server or internet hiccups at an unknown stage of the pipeline. We are unable to determine the exact cause for this failure at the moment although a thorough audit of the pipeline logs could potentially reveal the exact position and stage of the server or internet hiccups.

Overall, the results are satisfactory under our current conditions and resources used for load balancing. As detailed in Section II, our designs and cloud-based implementations are scalable and can be adapted to handle more users, more internet traffic, or perhaps even more complicated machine learning tasks in the future.

\section{Qualitative Testing and Debugging}

In terms of pipeline testing that cannot be quantified numerically or graphically, we have done exhaustive interaction with the front-end of the webpage in order to spot any UI/UX errors, or general bugs that would inhibit the user from utilizing our program. In order to test the back-end POST and GET requests, we used tooling like Postman in order to query our APIs to ensure correct output was being returned. In order to test the efficacy of our model, a standard test bed of data was evaluated, as described in Section III. A critical evaluation of our pipeline is also provided. 

\section{Future Directions Of Stratus} 

The initial goal of our project devised at the time of the proposal was to answer the following questions: 1) can we recognize digitally drawn digits using a deep learning model trained on the cloud; and 2) can we train a deep neural network in a distributed fashion using Spark ML. Exploring both of these particular queries in this project served as an initial MVP in order to deliver a proof-of-concept. Since we found that both sub-goals are indeed possible, we would like to expand our project in three main ways: 1) improved features and associated scalability; 2) accurate prediction; and 3) deployment in a large-scale production environment.

In particular, scalability in this project is primarily delivered via the use of Spark ML to allow for distributed training, as well as Docker and Kubernetes to achieve the use of multiple workers within our two primary virtual machines upon which our application is deployed. However, as more features are added to the application, the  scalability of the application, to handle both the load and expected increase in traffic must also be more resilient. For instance, a primary feature of interest that we would like to implement is the use of a multi-digit MNIST model, to allow the user to directly draw a multi-digit number, rather the just a single digit. The issue of scalability becomes more complex as now the size of the input into our model also scales with the length of the user input. Dealing with large input datasets into our website will present a challenge and will necessitate for a robust load handling protocol, similar to previous projects with respect to the energy counter dataset. Furthermore, leveraging more load balancing techniques as well as autoscaling would be nice to have in order to make the website more resilient than it currently is.  

Furthermore, another area of growth is improving the robustness of the pre-trained model that is leveraged within \textit{Stratus} for predictions. In particular, the current method in which our product works is that it fetches the user input and then down-samples it into an array of 28x28; such extreme down-sampling from the size of the canvas on the Document Object Model (DOM) to the input array of the model causes a loss of feature generality, and mitigates the salient distinguishing features of the user input that may distinguish it from other potential digits. A solution to this would be to preprocess our network, during the training phase, with input arrays that have been exposed to extreme down-sampling so as to allow for our model to perform more robustly on arrays with features that are not so clear. 

Finally, another area of growth would be to deploy our application in a production-like environment. While we did our best to emulate high stress and load to test our application, it would not truly match the experience of having several users submitting drawings to our websites, and seeing how our end-to-end pipeline manages multiple requests at once to our cloud infrastructure. Seeing the behavior of our product, specifically due to the deep learning component, which can suffer from large inference times, would be valuable information to have and would be a much-needed next step in our project. 

\section{Conclusion}

\textit{Stratus} is a proof-of-concept which leverages distributed cloud techniques and tools such as Spark ML, Apache Kafka, Chameleon Cloud, Ansible, Vagrant, Python Flask, Docker, and Kubernetes to implement a deep learning application. We have developed a holistic end-to-end pipeline by which a user is able to input a handwritten digit via a full stack application. The encoded drawing is sent to the cloud-infrastructure where a deep learning model predicts the number that the user drew and returns the prediction back to the user. We found success utilizing distributed and auto-scaling concepts via Spark, K8s, and NGINX.  Future directions of this work include: 1) improving scalability, 2) increasing robustness; and 3) real-world deployment. Overall, the novelty in this work is in leveraging the existing technologies used in class to create a real-world, production-grade full-stack, end-to-end application. 

\section*{Acknowledgment}
We would like to thank Dr. Aniruddha Gokhale (Vanderbilt University) for helping us learn more about cloud infrastructure at scale, as well as distributed systems concepts, which helped us conclude this project.

This group project was completed as part of Immersion Vanderbilt, a degree requirement for all undergraduate students at Vandebrilt University. Full details of the requirement can be found here: \href{https://www.vanderbilt.edu/immersion/}
{https://www.vanderbilt.edu/immersion/}. The project was demoed as part of Immersion Vanderbilt showcase at the end of Fall 2021 and the complete submission including the technical report and the complete source code was accepted by the Office of Immersion Resources on Apr 29, 2022 after an extensive review by a committee of faculty members.

\newpage
\section*{Appendix A}
In this appendix, we include the full GitHub repository that contains all the files we used to construct and deploy the project with detailed explanation for reproduction of both the server deployment within Chameleon Cloud and our demonstrated results. The GitHub repository is publicly available at \href{https://github.com/ruidazeng/stratus}
{\textit{https://github.com/ruidazeng/stratus}} under the MIT License. The original GitHub repository used during development can also be found at \href{https://github.com/aadarshjha/stratus}
{\textit{https://github.com/aadarshjha/stratus}}, it contains almost identical content, although this repository does not contain explicit licensing and is therefore under exclusive copyright by default.

Each deployment folder contains server configuration files in YAML format and the dockerfile used by Docker to build the images. The configuration files include the specifications and the ports and protocol to facilitate the service. These deployment folders contain most of the individual files needed to setup consumer, CouchDB, Flask, Kafka broker and zookeeper, NGINX, and Spark ML. These YAML files are included and referenced in the \path{playbook_demo_master.yml}, our master playbook used to execute our plays in order: (1) install packages on local VMs; (2) create chameleon VMs; (3) install packages on Chameleon VMs; (4) configure cloud files; (5) installing docker and kubernetes; (6) start kubernetes and taint master on VM2; (7) start kubernetes on VM3; (8) start docker image and create registry; (10) run Spark pipeline; (11) run NGINX pipeline; (12) run consumer.

The Python file \path{producerVM1.py} are used alongside shellcodes \path{iter_script.sh} and \path{bootstrap.sh} to deploy NGINX, Spark ML, and Kafka broker and zookeeper. The Python file \path{consumer.py} serves as the consumer in our machine learning pipeline within our cloud computing model. The \path{MyInventory}, \path{Vagrantfile}, and \path{ansible.cfg} are straightforward configuration files required to deploy the virtual machines on Chameleon Cloud. 

Last but not least, Python file \path{locustfile.py} contains source code we used to test both our load balancing quality for both the website hosting and for the prediction algorithm in Section III.

For further information and clarification regading the source code and reproduction of our results, the authors can be reached via the email addresses or via Issues/Pull request on GitHub.

\section*{Appendix B}
In this appendix, we include Figure 6-20. All of the figures are referenced and addressed in Section III. Evaluation and contain our quantitative test results for the load balancing efficiency and efficacy. We defined our user behavior using Python codes, with GET request used to test website hosting, and POST request used to test prediction algorithm. We then used the open source load testing tool LOCUST (\href{https://locust.io/}
{\textit{https://locust.io/}}) to swarm our system with the defined numbers of simultaneous users to utilitized the interface provided to generate the results represented by graphs.

\begin{figure*}[htbp]
\centerline{\includegraphics[width=200mm]{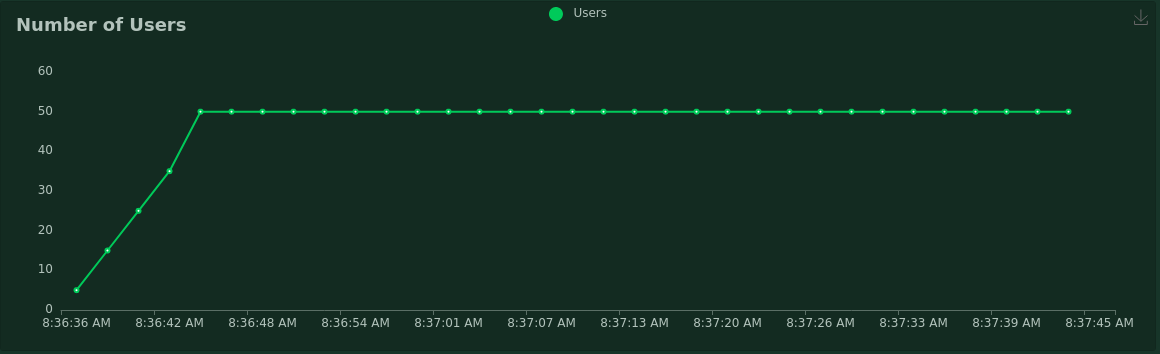}}
\caption{Number of Users with respect to Time for 50 Users.}
\label{fig}
\end{figure*}

\begin{figure*}[htbp]
\centerline{\includegraphics[width=200mm]{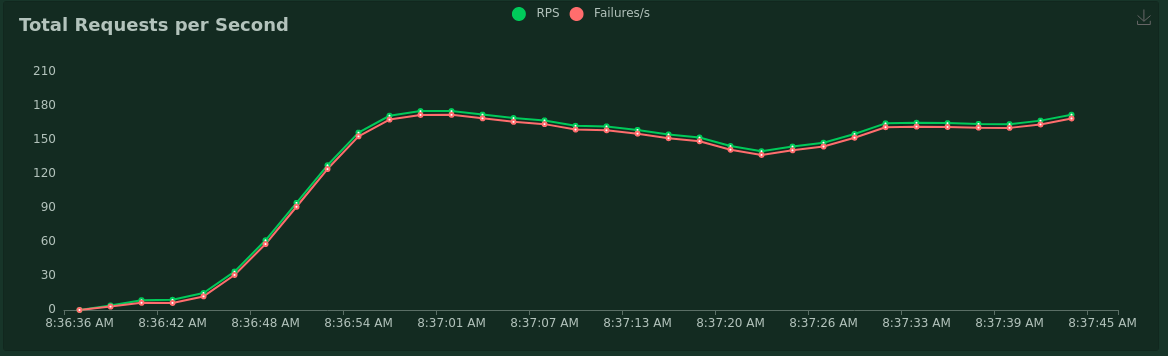}}
\caption{Total Requests Per Second for 50 Users.}
\label{fig}
\end{figure*}

\begin{figure*}[htbp]
\centerline{\includegraphics[width=200mm]{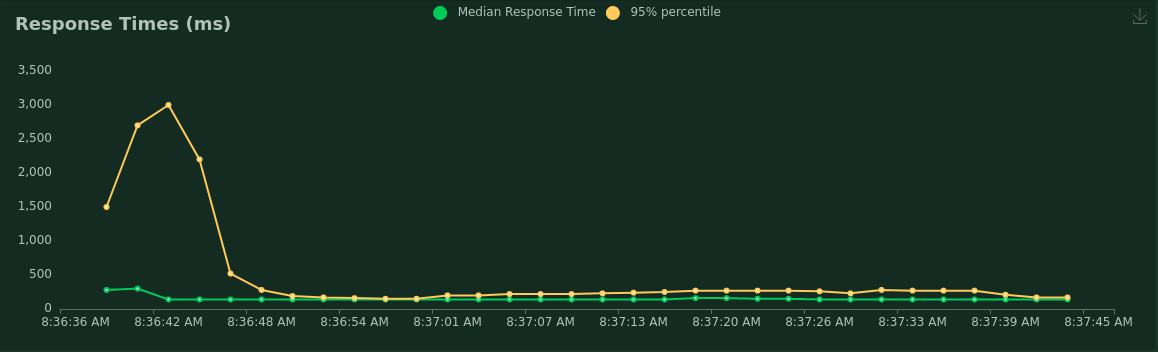}}
\caption{Response Times for 50 Users.}
\label{fig}
\end{figure*}

\begin{figure*}[htbp]
\centerline{\includegraphics[width=200mm]{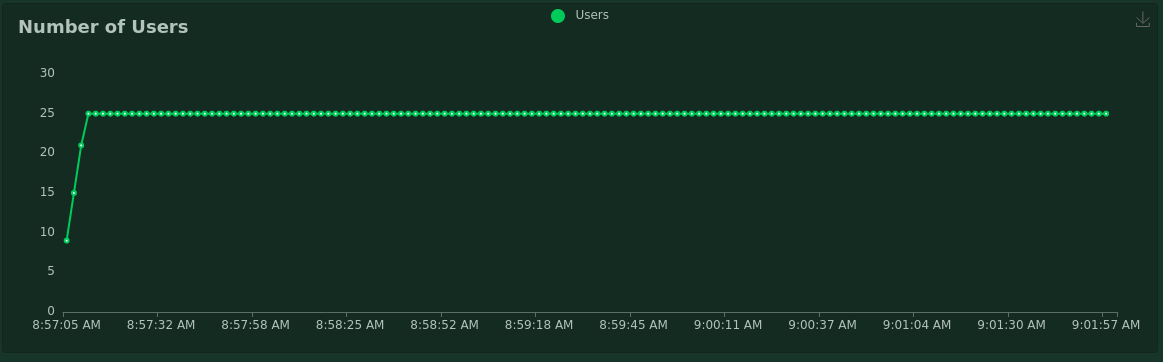}}
\caption{Number of Users with respect to Time for 25 Users.}
\label{fig}
\end{figure*}

\begin{figure*}[htbp]
\centerline{\includegraphics[width=200mm]{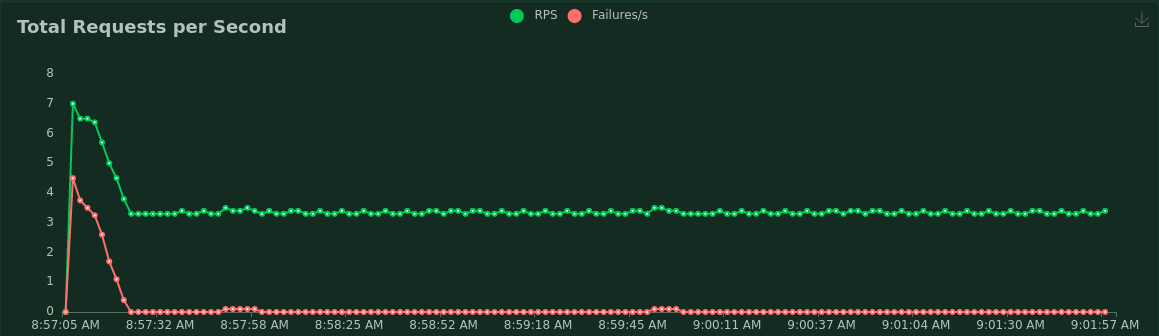}}
\caption{Total GET Requests Per Second for 25 Users.}
\label{fig}
\end{figure*}

\begin{figure*}[htbp]
\centerline{\includegraphics[width=200mm]{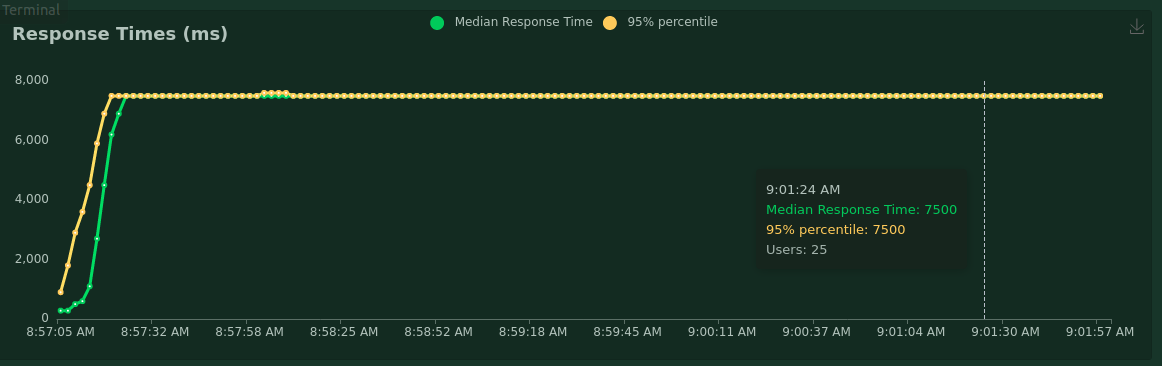}}
\caption{Response Times for 25 Users.}
\label{fig}
\end{figure*}

\begin{figure*}[htbp]
\centerline{\includegraphics[width=200mm]{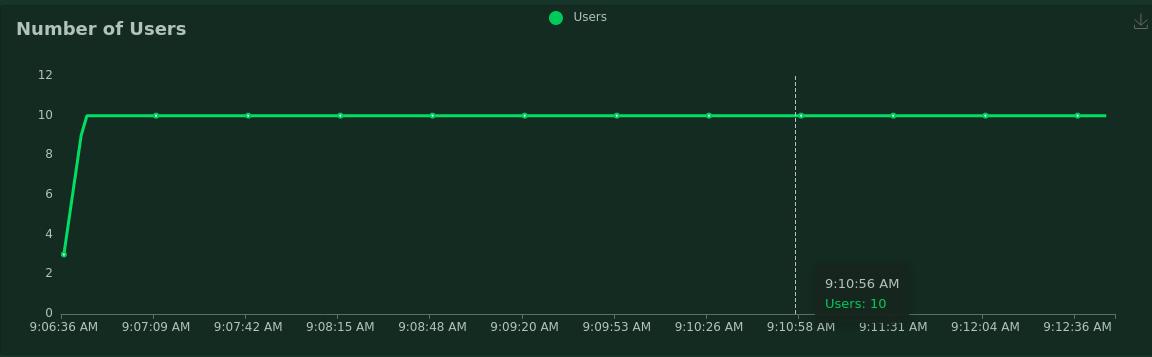}}
\caption{Number of Users with respect to Time for 10 Users.}
\label{fig}
\end{figure*}

\begin{figure*}[htbp]
\centerline{\includegraphics[width=200mm]{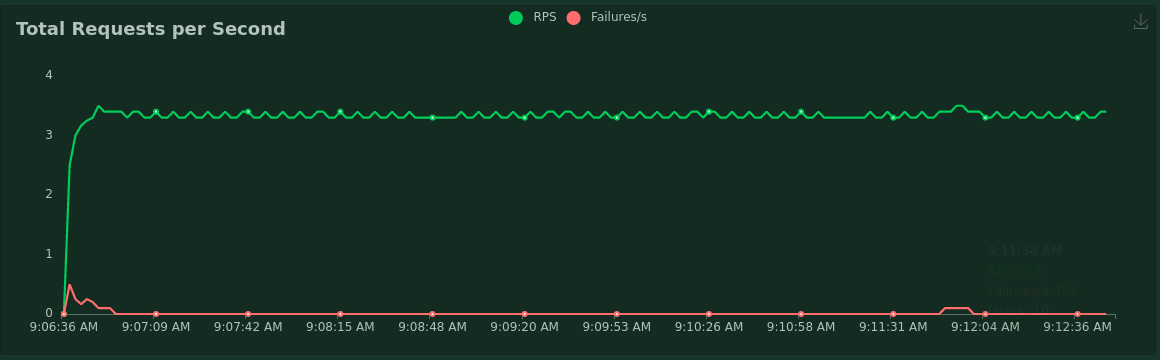}}
\caption{Total GET Requests Per Second for 10 Users.}
\label{fig}
\end{figure*}

\begin{figure*}[htbp]
\centerline{\includegraphics[width=200mm]{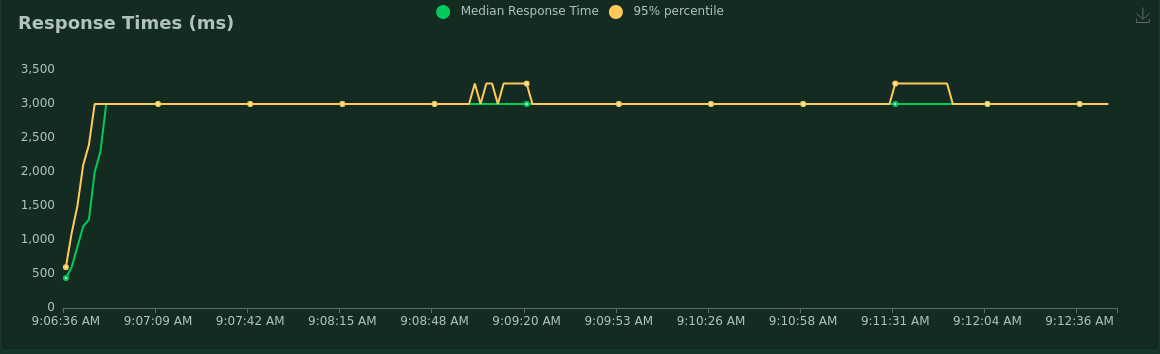}}
\caption{Response Times for 10 Users.}
\label{fig}
\end{figure*}

\begin{figure*}[htbp]
\centerline{\includegraphics[width=190mm]{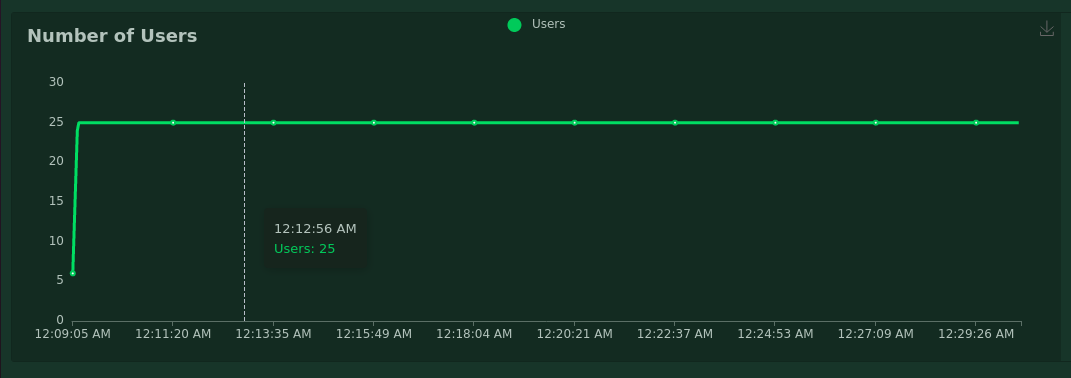}}
\caption{Number of Users with respect to Time for 25 Users.}
\label{fig}
\end{figure*}

\begin{figure*}[htbp]
\centerline{\includegraphics[width=190mm]{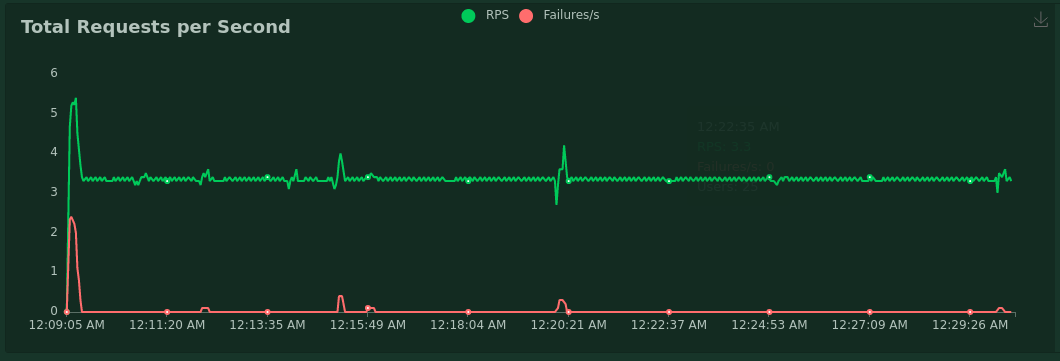}}
\caption{Total POST Requests Per Second for 25 Users.}
\label{fig}
\end{figure*}

\begin{figure*}[htbp]
\centerline{\includegraphics[width=190mm]{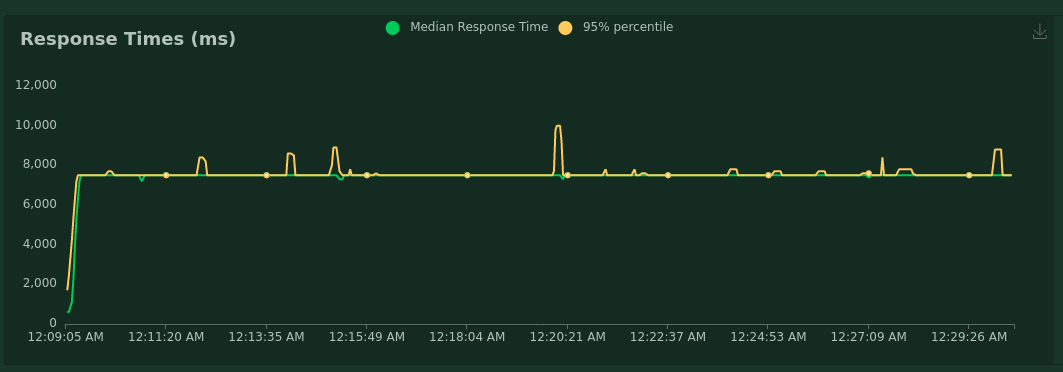}}
\caption{Response Times for 25 Users.}
\label{fig}
\end{figure*}

\begin{figure*}[htbp]
\centerline{\includegraphics[width=200mm]{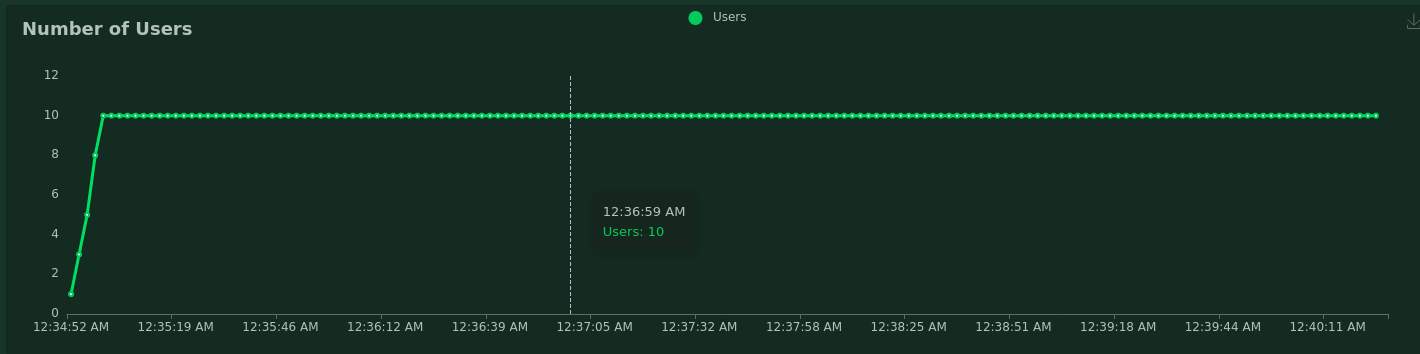}}
\caption{Number of Users with respect to Time for 10 Users.}
\label{fig}
\end{figure*}

\begin{figure*}[htbp]
\centerline{\includegraphics[width=200mm]{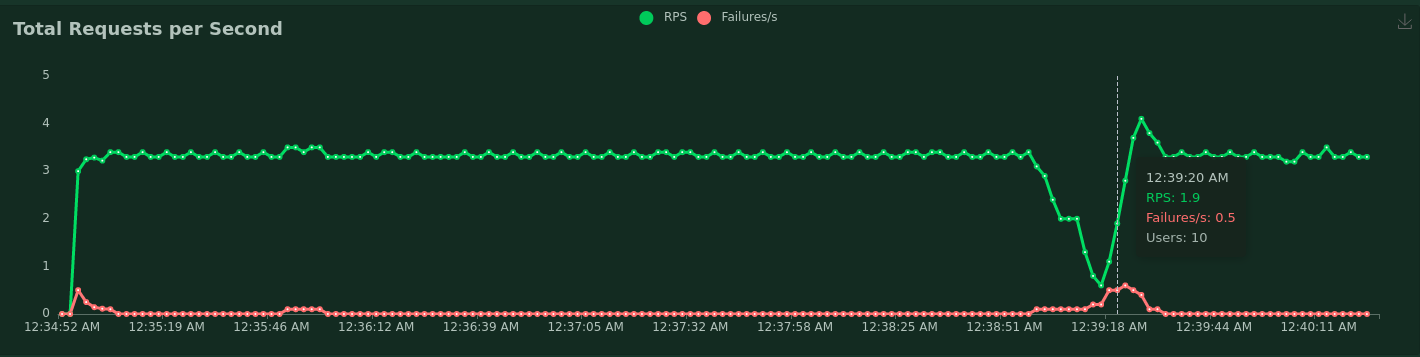}}
\caption{Total POST Requests Per Second for 10 Users.}
\label{fig}
\end{figure*}

\begin{figure*}[htbp]
\centerline{\includegraphics[width=200mm]{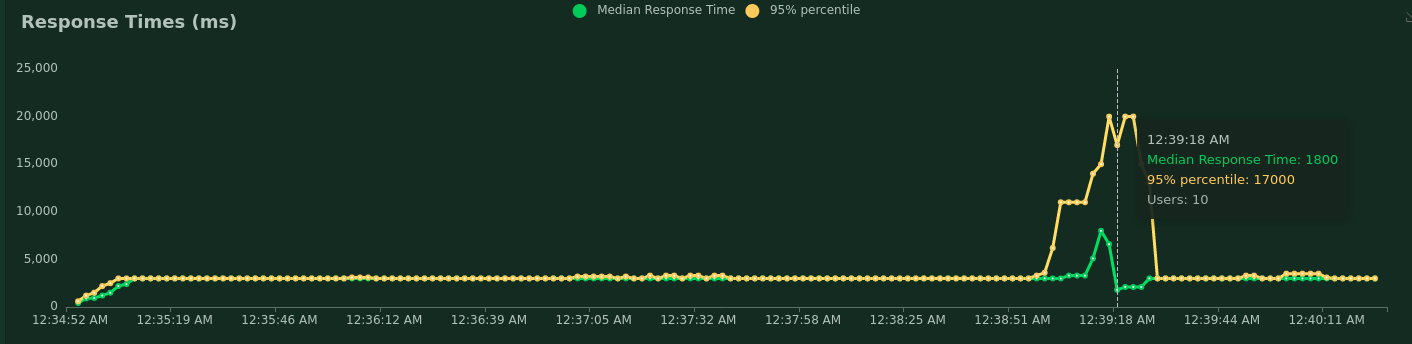}}
\caption{Response Times for 10 Users.}
\label{fig}
\end{figure*}

\end{document}